\documentclass[aps,floatfix]{revtex4}
\usepackage{epsfig}
\usepackage{color}
\usepackage{amssymb}
\usepackage{amsmath}
\usepackage[ngerman,english]{babel}
\usepackage{multirow}
\usepackage{graphicx}
\usepackage{subfigure}

\newcommand{\etal}{{\it et al.}}
\newcommand{\pr}[4]{Phys. Rev. #1 {\bf #2}, #3 (#4)}
\newcommand{\hedp}[3]{High Energy Density Phys. {\bf #1}, #2 (#3)}
\newcommand{\astropj}[3]{Astrophys. J. {\bf #1}, #2 (#3)}

\newcommand{\physfluid}[3]{Phys. Fluids {\bf #1}, #2 (#3)}

\newcommand{\rmnumb}[2]{{#1}_{\rm #2}}
\newcommand{\rmupper}[3]{{#1}^{{\rm #2}}_{#3}}

\newcommand{\bks}[1]{\left( #1 \right)}

\newcommand{\impart}[0]{{\rm Im}\, }
\newcommand{\repart}[0]{{\rm Re}\, }

\begin{document}

\title{Ionization potential depression and Pauli blocking in degenerate plasmas at extreme densities}

\author{Gerd R\"opke$^{1,2}$, David Blaschke$^{2,3,4}$, Tilo D\"oppner$^5$, Chengliang Lin$^1$, Wolf-Dietrich Kraeft$^1$, Ronald Redmer$^1$, and Heidi Reinholz$^{1,6}$}
\affiliation{
$^1$Institut f\"ur Physik, Universit\"at Rostock, D-18051 Rostock, Germany\\
$^2$National Research Nuclear University (MEPhI), 115409 Moscow, Russia\\
$^3$Institute of Theoretical Physics, University of Wroclaw, Wroclaw, Poland\\
$^4$Joint Institute for Nuclear Research, 141980 Dubna, Russia\\
$^5$Lawrence Livermore National Laboratory, Livermore, CA 94550, USA\\
$^6$University of Western Australia School of Physics, WA 6009 Crawley, Australia}
\date{\today}

\begin{abstract}

New facilities explore warm dense matter  (WDM) at extreme conditions where the densities are very high 
(e.g., carbon up to density of 50 g cm$^{-3}$) so that  electrons are degenerate even at 100 eV temperature.
Whereas in the non-degenerate region correlation effects such as Debye screening and its improvements are relevant for the ionization potential depression (IPD), new effects have to be considered in degenerate plasmas.
In addition to the Fock shift of the self-energies, the bound-state Pauli blocking becomes important with increasing density.
Taking these degeneracy effects into account leads to a reduction of the ionization potential and to a  higher degree of ionization.
Standard approaches to IPD such as Stewart-Pyatt and widely used opacity tables (e.g., OPAL)
do not contain Pauli blocking effects for bound states
so that they fail to explain experiments with WDM in the high density region. As example, results for the ionization degree of carbon plasmas are presented.

\end{abstract}

\maketitle

\section{Introduction}

The availability of new experimental facilities allows to explore 
matter under warm dense matter (WDM) conditions \cite{WarmDM},
where strong correlations in the ionic system and degeneracy 
of the electron system are of relevance.
The region of densities and temperatures that can be
probed has been extended towards multi-megabar pressures and 
temperatures up to tens of eV at synchrotrons, with pulsed power, 
high-power optical and free-electron-lasers or other methods of 
high-pressure experimental technique. There, strong correlations 
and quantum effects have to be treated consistently, and simple models
and approximations are pushed beyond their applicability limits. 

Within the model of the partially ionized plasma, WDM
consists of free electrons (particle density $n_e$) and ions $a_i$ with different 
ionization states $Z_i$ and densities $n_i$ (including the neutral atom with $Z_0=0$). 
It is characterized by the 
ionization degree $\bar Z = n_e/n_a$,  $n_a=\sum_i n_i$ being the particle density of all nuclei.
However, concepts such as the partially ionized plasma and the ionization degree have 
to be analyzed and applied with care,
because medium effects which influence the properties 
of isolated atoms and ions become more dominant with increasing density, leading to
shift  and broadening of energy levels and eventually 
to the disappearance of bound states (Mott effect, see Ref. \cite{Mottbuch}).
Nevertheless, the concept of the composition of a partially ionized plasma 
is a useful tool to investigate the consequences of the appearance of bound states
on thermodynamic properties, conductivity, optical spectra,
Thomson scattering spectra, and other physical properties.
However, near the Mott transition where the bound states merge with the continuum and are dissolved,
 the subdivision into (weakly) bound states and 
free states, including resonances, becomes questionable, 
and there exists no clear criterion to 
subdivide the electron subsystem into "free" and "bound" electrons. 
As discussed below, a many-body theory provides a consistent
approach to WDM allowing for a systematic treatment of correlations including bound state formation.

The properties of atoms and ions immersed in a dense plasma are modified owing to medium effects.
This refers also to the ionization energy  $I_i$ of the ion $a_i$ in the charge state $Z_i$, which at least is necessary to remove 
one electron from the ground state to the continuum of free electrons.
As a consequence, the ionization potential $I_i$ is modified compared to its vacuum value $I_i^{(0)}$.  
The ionization potential depression (IPD) $\Delta I_i=I_i^{(0)}-I_i$ is a
 particular property of WDM presently under intense discussion. 
In the low-density, weakly coupled limit, the shift of the energy of charged particles is given by screening. For an ion (atom) with charge state $Z_i$,
the well-known Debye result for the energy shifts (see, e.g., Refs. \cite{KKER,KSK}) leads to a reduction of the ionization potential
$I_i^{\rm Debye}=I_i^{(0)}-\Delta I_i^{\rm Debye}$ as compared to the unperturbed ionization energy $I_i^{(0)}$. For the global ionization process $a_i\rightleftharpoons a_{i+1}+e$ (further particles must  participate to realize conservation laws), we find the IPD in Debye approximation
\begin{equation}
\label{Debye}
\Delta I^{\rm Debye}_i=\kappa_{\rm class} (Z_i+1) \frac{e^2}{4 \pi \epsilon_0},\qquad \kappa^2_{\rm class}=\frac{e^2}{\epsilon_0 k_BT} \left(\sum_i Z_i^2n_i+n_e\right)\,.
\end{equation}
A more general expression for the screening parameter (inverse Debye radius) $\kappa=1/r_D$ which takes into account also the degeneracy of electrons, is given below in Sec. \ref{inmedSG}.

At high densities where the ions are strongly correlated, the ion sphere model, see \cite{IS}, is more adequate.
In contrast to the Debye approximation, the density dependence of the shift is weaker ($\propto n^{1/3}$).
Semi-empirical interpolations have been proposed by Ecker and Kr{\"o}ll (EK) \cite{EK63} and Stewart and Pyatt (SP) \cite{SP66}
which are frequently used for estimating the IPD. In the SP approach, the IPD is given according to
\begin{equation}
\label{SP}
 \Delta I^{\rm SP}_i=\frac{3}{2} \frac{(Z_i+1) e^2}{4 \pi \epsilon_0 r_{\rm IS}} 
\left[(1+s^3)^{2/3}-s^2 \right]
\end{equation}
with the ion sphere radius $r_{\rm IS}= (3 Z_i/(4 \pi n_e))^{1/3}$ and $s=(\kappa r_{\rm IS})^{-1}$.
New experiments on high density plasmas 
\cite{Hoarty13,Vinko12,VCW14,Vinko15,Vinko18,ciricosta12,ciricosta16} cannot be explained using any of these simple 
approximations, and the need for a better approach is obvious
when going to extreme conditions where the ions are strongly coupled.

According to Crowley  \cite{Crowley14},  the chemical picture has to be replaced by a physical picture 
based on quantum statistical many-body theory \cite{KKER,KSK}.
In this work, such a systematic treatment of different plasma effects is worked out using Green-function techniques.
Alternatively, numerical simulations such as path integral Monte-Carlo (PIMC) simulations \cite{Militzer,Dornheim} 
have been used to give a systematic approach to properties of WDM.
Because of the fermion sign problem, PIMC simulations of two-component plasmas are restricted
presently to high temperatures and high densities. 

The electron-ion interaction is strong in the low-temperature region where bound state
formation is relevant. A very successful and practicable approximation is
density-functional theory (DFT) for electronic structure calculations 
in combination with molecular dynamics simulations for the ions (DFT-MD).
It has proven to predict results for WDM states, see,
e.g., Ref. \cite{DFT}. Using  standard expressions for the
exchange-correlation part of the (free) energy,  
detailed properties of the electron system like the density of states 
as well as the ionic structure factor are obtained. Electron-electron correlations are
treated approximately using appropriate expressions for the energy-density functional. For the treatment of IPD using this formalism see 
Refs. \cite{Vinko12,VCW14,Vinko15,Vinko18}.

Coming back to the quantum statistical approach
using Green-function techniques, the 
shift of the continuum is related to the single-particle self-energy.
A systematic discussion of the energy spectrum of hydrogen atoms in dense plasmas
within the Green function approach has been given by Seidel {\it et al.} \cite{Seidel}.
An improved treatment of the self-energy using the Montroll-Ward 
expression, which gives the Debye shift  in the low-density 
limit,  has been proposed recently by Lin {\it et al.}  \cite{Lin17}. 
Using the fluctuation-dissipation theorem,
the inverse dielectric function is related to the dynamical
structure factor. With known expressions for the ion-ion 
structure factor (SF), the calculated IPD show a better agreement with experimental data.

If going to even higher densities, in addition to the strong coupling of the ions, the degeneracy of
electrons becomes important. Related effects, in particular Pauli blocking and Fock shifts, 
are not included
in the approaches to the IPD \cite{SP66,Lin17} discussed so far. 
The electron degeneracy parameter is defined as 
\begin{equation}
\Theta = \frac{T}{T_{\rm Fermi}}=\frac{2 m_e k_BT}{\hbar^2} (3 \pi^2 n_e)^{-2/3}\,.
\end{equation}
In the region of the temperature-density plane, where $\Theta \le 1$,  a classical description 
is no longer valid. Instead, quantum effects, in particular the Pauli principle as a consequence 
of the antisymmetry of the many-electron (fermonic) wave function, lead to so-called exchange terms. 
The related condition
$n_e \Lambda_e^3 \ge 1$ with 
$n_e \Lambda_e^3 = n_e (2 \pi \hbar^2/m_e k_BT)^{3/2}=8/(3 \pi^{1/2}) \Theta^{-3/2}$
is well known as a condition where the classical gas approach is not applicable, 
and the quantum description based on the Fermi distribution function must be applied. 
For instance, in carbon plasmas at $T=100$ eV, the electrons become degenerate at electron density 
$n_e \approx4 \times 10^{24}$ cm$^{-3}$ corresponding to a carbon mass density of 20 g cm$^{-3}$.
New experiments are planned and will be performed, for instance, at the National Ignition Facility (NIF)
in Livermore to explore WDM  \cite{WDM} at very high densities where plasmas become degenerate even 
at temperatures of the order of 100 eV. 

New physics becomes of importance in degenerate systems. Whereas at lower densities, in the classical 
region, dynamical screening is the most important medium effect, at  extremal high densities exchange 
effects become of increasing relevance.
Note that opacity  tables (OPAL) using the SP approximation for the IPD are available \cite{OPAL} and 
frequently used. However, 
degeneracy effects such as bound-state Pauli blocking and Fock shifts are not consistently included.
Pauli blocking effects have been extensively investigated for light clusters ($^2$H, $^3$H, $^3$He, $^4$He) 
in nuclear matter \cite{R}, see also Ref. 
 \cite{Ropke:1986qs} as a mechanism of hadron dissociation.
For hydrogen plasmas they have been discussed in Ref.  \cite{EBRRR09}.

In this work we give a systematic treatment of the effects of degeneracy within a Green function approach 
and its consequences for IPD in the region of very high densities where standard approaches such as SP 
used, e.g., for OPAL, become inapplicable.
In the following Sec. \ref{sec:SE} we consider the effective wave equation for few-particle (bound) states in a plasma environment
and discuss Pauli blocking and Fock shifts.
We apply these results to carbon plasmas at high densities in Sec. \ref{sec:results},
where a significant increase of the ionization degree compared to the 
frequently used SP model is obtained.
Pauli blocking effects are also of relevance for  K-edge shifting \cite{Hu} to be discussed in the conclusions.

\section{In-medium Schr\"odinger Equation}
\label{sec:SE}

\subsection{Low-density limit of the plasma composition}

We consider an element $a$ (e.g., carbon C in Sec. \ref{sec:results}) in the WDM region.
Let us first recall the low-density limit where the in-medium effects can be neglected. 
The model of partially-ionized plasma (PIP) considers a plasma which is composed of different ions ($a_i$) with charge number $Z_i$
at partial density $n_i$ (including neutral atoms), as well as free electrons at density $n_e$.
The components of the PIP can react, changing the state of excitation, including ionization and recombination processes. Thermodynamic equilibrium is described by relations between the corresponding chemical potentials of the different components.

The ideal electron chemical potential for arbitrary degeneracy follows from the expression 
for the density
\begin{equation}
\label{nemu}
 n_e= g_e \int \frac{d^3p}{(2 \pi)^3} \frac{1}{\exp[\beta\hbar^2 p^2/(2 m_e)-\beta\mu_e]+ 1}
\end{equation}
 where $\beta=1/(k_BT)$. The factor $g_e=2$ accounts for spin degeneracy. The Fermi distribution 
 can be replaced by the Boltzmann distribution in the classical case $\exp(\beta \mu_e)\ll 1$ so that
$\exp(\beta \mu_e)\approx n_e (2 \pi \beta \hbar^2 /m_e)^{3/2} /2$.

The density $n_{i, \hat n}$ of ions with its fully known quantum state $\hat n$, characterized by the 
complete set of quantum numbers including total momentum, spin, angular momentum, etc., 
is given by the chemical potential $\mu_ {i, \hat n}$. For instance, for the two-body problem, the complete set of quantum numbers ${\hat n}=\{{\bf P}, \gamma, \nu\}$ contains in addition to the center-of mass momentum $\bf P$ and the channel quantum number $\gamma$ further intrinsic quantum numbers $\nu$ which describe the intrinsic excitation. The channel quantum number $\gamma$ contains, e.g., spin and angular momentum depending on the observables which are conserved in the two-body interaction. 

In the low-density limit, where the interaction between the particles and clusters
can be neglected (with exception of reacting collisions), the summation over the total momentum $\bf P$
can be performed, and we obtain in thermodynamic equilibrium the well-known relation 
for the ideal gas 
\begin{equation}
n_ {i,\gamma, \nu}=\int \frac{d^3 P}{(2 \pi)^3}e^{-\beta \hbar^2P^2/(2M)+\beta \mu_{i,\gamma, \nu}}=
\frac{1}{ \Lambda^{3}} e^{\beta \mu_{i,\gamma, \nu}},
 \end{equation}
where $\Lambda = [2 \pi \beta \hbar^2/M]^{1/2}$
is the  thermal wavelength of the ions. We restrict us to the region of thermodynamic parameters where 
the ions can be treated classically,  $M$ is the ion mass (dependence on charge number is neglected).
The specific chemical potentials $ \mu_{i,\gamma, \nu}$ are gauged so that only the kinetic energy 
of the cluster owing to the center-of-mass motion is considered. The potential energy, 
in particular the binding energies of ions, must be considered separately. 

The description is simplified if we consider also the sum over intrinsic degrees of freedom, similar to the spin degeneracy in the case of the electron component. With respect to the ground state $\{\gamma, \nu\}=(0)$ of the ion $a_i$ with the chemical potential $\mu_i$ for this ground state, the excitation energy is denoted by $E_{i,\gamma, \nu}$ (we assume that the energy of the intrinsic motion does not depend on $\bf P$). Chemical equilibrium is achieved if the condition
$\mu_{i,\gamma, \nu}=\mu_i+ E_{i,\gamma, \nu}$ holds. For the total contribution of ions $a_i$
with charge $Z_ie$ we have
\begin{equation}
\label{ni1}
 n_i=\sum_{\gamma, \nu}n_{i,\gamma, \nu}=\frac{1}{ \Lambda^{3}}\sum_{\gamma,\nu}
e^{\beta (\mu_i+ E_{i,\gamma, \nu})}=\frac{1}{ \Lambda^{3}}\sum_\gamma \sigma_{i,\gamma}(T)e^{\beta \mu_i}
\end{equation}
with the intrinsic partition function in the channel $\gamma$
\begin{equation}
 \sigma_{i,\gamma}(T)=\sum_{\nu} e^{\beta E_{i,\gamma, \nu}}.
\end{equation}
In a further step, we can also perform the sum over the different channels to obtain the full intrinsic partition function $\sigma_i(T) = \sum_\gamma  \sigma_{i,\gamma}(T)$ of the ion $a_i$ so that 
\begin{equation}
\label{compni}
 n_i=\frac{1}{ \Lambda^{3}} \sigma_{i}(T)e^{\beta \mu_i}\,.
\end{equation}

Compared to the expression for the free electrons (\ref{nemu}), the spin summation is contained in the summation over $\gamma$, and the intrinsic excitations are taken into account because the ion $a_i$ is in general a composed particle.
An important issue is that the summation over $\nu$ in Eq. (\ref{ni1}) has to be performed not only over the bound states 
but also over the continuum of scattering states, where the quantum number $\nu$ is replaced by the energy 
$E$ of relative motion and $\delta_{i, \gamma}(E)$ is the scattering phase shift in the channel $\gamma$.
According to Beth and Uhlenbeck \cite{BU}, the following expression for the contribution 
of correlations from the channel $\gamma$ 
of the intrinsic partition function can be derived:
\begin{eqnarray}
\label{B-U}
&&\sigma_{i, \gamma}(T)=\sum_\nu^{\rm bound}   e^{-\beta E_{i, \gamma, \nu}}+\int_0^\infty \frac{dE}{\pi} 
e^{-\beta E}\frac{d}{d E} \delta_{i, \gamma}(E) 
\end{eqnarray} 
or, after integration by parts and using the Levinson theorem \cite{Bolle,Girardeau}, 
\begin{eqnarray}
\label{n2virial}
&&\sigma_{i, \gamma}(T)= 
 \sum_\nu^{\rm bound}   \left(e^{-\beta E_{i, \gamma,\nu}}-1\right)+\int_0^\infty \frac{dE}{\pi k_BT} 
e^{-\beta E} \delta_{i, \gamma}(E). 
\end{eqnarray}
In particular, calculating the pressure as function of density and temperature by integration 
of $n(T,\mu_i,\mu_e)$, the
results (\ref{B-U}), (\ref{n2virial}) give exact expressions for the second virial coefficient \cite{BU}.
 
At this point, an important comment is necessary. The subdivision into the contribution of bound states 
and scattering states
 is model-dependent. In particular, there is no 
clear criterion to define bound states near the continuum edge, because there is no principal 
difference between the physical properties of a weakly bound state and a resonance state in the continuum. 
Eqs. (\ref{B-U}) and (\ref{n2virial}) are different 
with respect to the subdivision into a discrete part related to the bound states and the continuum part. 
Therefore, the  definition of the ionization degree $\bar Z=n_e/n_a$
is model dependent. Physical properties such as the second virial coefficient, should take into account 
also the contribution of scattering states and are independent of the subdivision into bound and 
scattering state contributions. 
To define the ionization degree, we use the correlated part of the second virial coefficient 
which remains after subtraction of all 
single quasiparticle contributions, see Eq. (\ref{n2virialqu}) in the following section. 
In particular, this correlation part contains in addition to 
contributions of the bound states also a contribution owing to resonances if they exist.

In the case of Coulomb interaction, scattering phase shifts can not be defined in the standard way because of the long-range character 
of the Coulomb potential. This problem has been investigated in plasma physics since a long time, see Refs. \cite{OPAL,KKER,KSK,Rogers}. 
We give here a result for the intrinsic partition function of hydrogen-like ions, 
the Planck-Larkin expression 
\begin{equation}
\label{PLC5}
\sigma^{\rm Planck-Larkin}(T)= 
\sum_{n=1}^{\infty} 2n^2 \left[e^{-\beta E_{n}}-1+\beta E_{n}\right]
\end{equation}
where $n$ runs over the intrinsic quantum numbers (including spin) of all bound states, 
$E_n=-Z^2e^2/(4 \pi \epsilon_0 a_B n^2)$.  

Reactions in partially ionized plasmas (PIP) include also ionization and recombination processes
$ a_{i,\gamma, \nu} +s\rightleftharpoons a_{i+1,\gamma', \nu'}+e+s'$.
The conservation laws (in particular of energy and momentum) demand, for instance, 
the collision with a third particle $s$ (spectator) or the emission/absorption of a photon.  
In thermodynamical equilibrium, these processes lead to 
a relation between the chemical potentials $\mu_i$  introduced above for the ground state of the corresponding ionic components of the PIP. Using the notation $I_i$ (ionization potential) for the lowest 
excitation energy $E_{i,0}$ of the ionic ground state to become ionized, 
the condition for chemical equilibrium reads
\begin{equation}
\label{MAL}
\mu_i=\mu_{i+1}+\mu_e+I_i\,.
\end{equation}
The bound state energy $-I_i$ 
(ground state energy of the ion $a_i$ relative to the continuum of $a_{i+1}+e$) can be 
implemented as potential energy in the scaling of the chemical potentials of each ion charge state. 
Inserting relation (\ref{MAL}) in Eq. (\ref{compni}), the Saha equation is obtained which determines 
the concentration of the different components of the PIP. If there are several ionization states $Z_i$, 
the repeated use of Eq. (\ref{MAL}) leads to a coupled system of Saha equations.
Finally, only the chemical potentials of the electrons and the ionic nuclei remain, corresponding 
to the conserved total number of electrons and nuclei of the WDM. 
Taking into account electrical neutrality, the thermodynamic state of WDM is defined by the total mass density
$n^{\rm total}_a$ and the temperature $T$. The composition of the PIP model, 
including the degree of ionization $\bar Z$,
follows from the solution of the coupled system of Saha equations (\ref{compni}), (\ref{MAL}).
Results for the composition within the PIP model in the low density region, 
neglecting the interaction between the components, are well known.
According to the mass-action law, the ionization degree $\bar Z$ increases with increasing $T$, 
but decreases with increasing $n^{\rm total}_a$. 
Results for the ionization degree of the ideal carbon plasma 
are given below in Fig. \ref{fig:2a}.

\subsection{Quantum statistical approach for interacting plasmas}
\label{QS}

The definition 
of the composition of a dense system is not free of model assumptions so that one should use a 
systematic quantum statistical approach to calculate physical properties. 
Nevertheless, the composition of a PIP and a corresponding ionization degree 
are useful concepts for low-density plasmas, but have to be handled with care in the high-density region. 
In the present work, we consider thermodynamics
to define the composition and the ionization degree of the PIP. 
Instead of the contribution of free electrons to the total density, free quasiparticles with 
medium-dependent energies are considered. 
The remaining part of composition describes correlations, in particular the contribution of bound states.
A quantum statistical approach to the composition of a PIP is obtained from the 
equation of state which relates the total densities of electrons $n_e^{\rm total}$ 
and nuclei $n_a^{\rm total}$ to the temperature $T=1/(k_B \beta)$ and the chemical 
potentials $\mu_e, \mu_a$,
\begin{equation}
\label{specn}
 n^{\rm total}_e(T,\mu_e,\mu_a)= \frac{1}{\Omega}\sum_1 \int_{\infty}^\infty \frac{d \omega}{2 \pi}
\frac{1}{e^{ \beta (\omega-\mu_e)}+1} A_e(1,\omega)
\end{equation}
with the spectral function $A_e(1,\omega)$, the single-particle states are denoted by wave number vector and spin, $|1\rangle=|{\bf p}_1,\sigma_1 \rangle$, and $\Omega$ the system volume. A corresponding relation holds also 
for $n_a^{\rm total}$.
Both the variables $\mu_e, \mu_a$ are related to each other because of charge neutrality. 
With the charge number $Z_a$ of the nuclei, 
we have $n_e^{\rm total}=Z_a\,n_a^{\rm total}$. 
The relation (\ref{specn}) gives an immediate access to the mass action law or, in plasma physics, the Saha equation. Having the equations of state $\mu_c(T,Z_a\,n^{\rm total}_a,n^{\rm total}_a)$, $c=a,e$, to our disposal, 
thermodynamic potentials such as the free energy $F(T,Z_a\,n^{\rm total}_a,n^{\rm total}_a)$
are obtained by integration. From this, all other thermodynamic properties are derived. Note that also the density of states is obtained from the spectral function.

The spectral function, which fulfills the normalization condition $\int \frac{d \omega}{2 \pi}
 A_e(1,\omega)=1$, is related to the self-energy
\begin{equation}
\label{spec1}
 A_e(1,\omega)= \frac{2\, {\rm Im} \Sigma_e(1,\omega+i0)}{[\omega-E_e(1)-{\rm Re} \Sigma_e(1,\omega)]^2+
[\Sigma_e(1,\omega+i0)]^2}\,.
\end{equation}
The self-energy, which is defined by the Dyson equation for the Green function as $G_e(1, iz_\nu)=
1/[iz_\nu-E_e(1)-\Sigma_e(1, iz_\nu)]$, can be calculated for given interaction using the technique of Feynman's diagrams.
For small ${\rm Im} \Sigma_e(1,\omega+i0)$, i.e. small damping of the quasiparticles, we have
\cite{ZS85}
\begin{equation}
\label{spec2}
 A_e(1,\omega)\approx \frac{2\pi \, \delta(\omega-E_e^{\rm quasi}(1))}{1-\frac{d}{dz} {\rm Re}
\Sigma_e(1,z)|_{z=E_e^{\rm quasi}-\mu_e}}-2 {\rm Im} \Sigma_e(1,\omega+i0) \frac{d}{d \omega}
\frac{{\cal P}}{\omega+\mu_e-E_e^{\rm quasi}(1))}
\end{equation}
with the quasiparticle energy 
\begin{equation}
\label{quasip}
 E_e^{\rm quasi}(1)=E_e(1)+{\rm Re} \Sigma_e(1,\omega)|_{\omega=E_e^{\rm quasi}}=E_e(1)+\Delta_e(1)
 \end{equation}
and $\cal P$ denoting the principal value. For the self-energy $\Sigma_e(1,z)$, a cluster decomposition 
can be performed, which leads to mass action laws \cite{R82}.

Considering two-particle contributions (T-matrix) to the self-energy,
we obtain the generalized Beth-Uhlenbeck formula for the virial expansion in the quasiparticle picture \cite{ZS85,SRS}
\begin{eqnarray}
\label{n2virialqu}
&&n^{\rm total}_e(T,\mu_e,\mu_a)= \frac{1}{\Omega}\sum_1 f_e(E^{\rm quasi}(1))\\&&+
\frac{1}{\Lambda^3}\sum_{i, \gamma}Z_i e^{\beta \mu_i}\left[\sum_{\nu}^{\rm bound} 
 (e^{- \beta E_{i, \gamma, \nu}}-1)+\frac{ \beta}{\pi} \int_0^\infty dE
e^{- \beta E}\left\{ \delta_{i, \gamma}(E)
 -\frac{1}{2} \sin [2  \delta_{i, \gamma}(E)]\right\} \right], \nonumber
\end{eqnarray}
$ f_e(E)=\{\exp[\beta (E-\mu_e)]+1\}^{-1} $, and $ E_{i, \gamma, \nu}$ is the excitation energy 
of the ion $a_i$, channel $\gamma$. The contribution of free electrons is replaced by the contribution of 
quasi-single particles with shifted energies (\ref{quasip}).
The contribution of the scattering states is reduced (sin-term in the last expression) 
because part of the interaction in the continuum, 
in particular the contribution of Born approximation, is already accounted for introducing the quasi-single particle contribution \cite{SRS,cvir}. As discussed in the following section \ref{inmedSG}, the bound state energies $E_{i, \gamma, \nu}$ and scattering phase shifts $\delta_{i, \gamma}(E)$ are modified by the interaction with the surrounding plasma as well and are calculated from an in-medium Schr{\"o}dinger equation.

At this point we can perform a subdivision of the total electron density into a free part given by the 
(damped) quasi-single particle contribution, and the remaining correlated density contribution. 
This definition of a free electron density $n_e$ and the corresponding ionization degree is possible 
as long as the single-electron spectral function (\ref{spec1}), (\ref{spec2}) shows a peak structure owing to the quasiparticle excitation.
Within a cluster decomposition of the self-energy, a similar decomposition can also be performed 
for the higher order T-matrix contributions, see the cluster-virial expansion discussed 
for nuclear matter in Ref.  \cite{cvir}. A cluster quasiparticle contribution (cluster mean-field
approximation) is discussed in the following section \ref{inmedSG}.

As a consequence, the cluster contributions $n_i$ of the ionization state $Z_i$ to the ion density and the 
 density of electrons is not restricted to only the bound state contribution, but contains also continuum contributions given in terms of the scattering phase shifts as shown in the second part of the right-hand side of Eq. (\ref{n2virialqu}).

\subsection{In-medium Schr{\"o}dinger equation and density effects}
\label{inmedSG}

The ideal plasma with the unperturbed energies $E_{i, \gamma, \nu}$ of the bound states and the kinetic 
energies of the free states cannot describe plasmas at high densities where interaction 
effects are important.
For simplicity we consider here the ionization degree of carbon at very high densities 
and/or temperatures where the carbon atoms are either fully ionized or in the C$^{5+}$ state, 
i.e. with one bound electron.

We consider the in-medium two-particle problem of the formation of the C$^{5+}$ state, described by the 
two-particle in-medium Schr\"odinger  equation. A Green function approach \cite{KKER,KSK}
 leads to the following two-particle equation (quantum number ${\hat n}=\{{\bf P},\gamma, \nu\}$, 
total momentum $\bf P$, spin variable not given explicitly)
\begin{eqnarray}
\label{BSE}
&&\left[E_e(p )+\Sigma_e(p,z)+E_{{\rm C}^{6+}}(k)+\Sigma_{{\rm C}^{6+}}(k,z)
\right] \psi^{5+}_{\hat n}({\bf p},{\bf k})\nonumber\\&&
+[1-f_e(p )\mp f_{{\rm C}^{6+}}(k)] \sum_{\bf q}V_{{\rm C}^{6+}, e}^{\rm eff}({\bf p},{\bf k},{\bf q},z) \psi^{5+}_{\hat n}({\bf p}+{\bf q},{\bf k}-{\bf q})=
E^{5+}_{\hat n} \psi^{5+}_{\hat n}({\bf p},{\bf k})
\end{eqnarray}
with the effective interaction
\begin{eqnarray}
\label{Veff}
&&V_{{\rm C}^{6+}, e}^{\rm eff}(1,2,{\bf q},z)= V_{{\rm C}^{6+}, e}(q)\left[1-\int_{-\infty}^\infty \frac{d \omega}{\pi}
{\rm Im}\, \varepsilon^{-1}(q,\omega+i0)\left[n_B(\omega)+1\right]\right.\nonumber \\&&
\left.\times\left[ \frac{1}{z-\omega-E_e(1)-E_e(2-{\bf q})}+\frac{1}{z-\omega-E_e(1+{\bf q})-E_e(2)}
\right]\right]\,.
\end{eqnarray}
We neglected higher order terms $\propto f_c( p)=\{\exp[\beta (E_c( p)-\mu_c)]\pm 1\}^{-1}$, the Fermi/Bose function for $c=e,{\rm C}^{6+}$. $V_{{\rm C}^{6+}, e}(q)=-Z_6 e^2/\epsilon_0 q^2$ is the Coulomb interaction. Without any chemical potential,  
$n_B(\omega)=[\exp(\beta \omega)-1]^{-1}$ is the Bose distribution function. 

The in-medium Schr{\"o}dinger equation (\ref{BSE}) contains the contribution of self-energies $\Sigma_c$ as well as 
the contribution of effective interaction including Pauli blocking. As a consequence, the energy 
eigenvalues $E^{5+}_{\hat n}$ of bound states as well as of continuum states are dependent on density and 
temperature of the surrounding plasma. These "dressed" states are denoted as quasiparticle excitations.
The Mott effect is the disappearance of a bound state 
if the ionization potential $I^{5+}_{\gamma, \nu}=E^{5+}_{\rm cont}-E^{5+}_{\gamma, \nu}$ 
goes to zero (Mott density). 
The continuum edge  $E^{5+}_{\rm cont}=\Delta_{{\rm C}^{6+}}(k=0)+\Delta_e(p=0)$ is given by the quasiparticle shifts (\ref{quasip}).
The bound state energy $E^{5+}_{\gamma, \nu}$ 
is a function of temperature and density. 
The bound state part of the intrinsic partition function (\ref{ni1}), (\ref{n2virial}), (\ref{n2virialqu}) has the form
\begin{equation}
\label{virC5}
\sigma^{\rm bound}_{C^{5+}}(T)= 
\sum^{\rm bound}_{\gamma, \nu} \left[e^{\beta I^{5+}_{\gamma, \nu}}-1\right] 
\theta\left(I^{5+}_{\gamma, \nu}\right),\qquad 
\theta(x)=\left\{\begin{array}{ll}
        1 &  \,\,{\rm if}\, x>0,\\
0& \,\, {\rm else},
       \end{array}
\right\}
\end{equation}
with the channel $\gamma$ (spin, angular momentum) and the intrinsic  excitation $\nu$. Later on we use this approximation for the generalized Beth-Uhlenbeck (BU) formula (\ref{n2virialqu}) 
to define the density contribution of bound states.
As a function of temperature and density, the intrinsic partition function (\ref{virC5}) is continuous at the Mott density. 
A more detailed approach based on in-medium scattering phase 
shifts, see last term in Eq. (\ref{n2virialqu}), can also take into account resonances in the continuum.

In the zero density limit where in-medium effects are absent, Eq. (\ref{BSE}) reproduces the Schr\"odinger equation for the 
hydrogen-like atom. Density effects arising from the dynamical self-energy $\Sigma_c(p,z)$, 
the Pauli blocking $(1-f_e\mp f_{{\rm C}^{6+}})$, and the dynamical screening expressed by the dielectric function $\varepsilon(q,z)$ in Eq. (\ref{Veff}) have to be treated in appropriate approximations.
As mentioned above, the carbon ions can be treated classically so that the contribution 
$\mp f_{{\rm C}^{6+}}(k)$ can be dropped. The Pauli blocking becomes relevant if the free electrons are degenerate. 
The contribution to the shift of bound state energies is discussed in the following section \ref{sec:deg}.

Using the technique of Feynman diagrams, systematic approaches for the dielectric function  $\varepsilon(q,z)$ can be found \cite{KKER}. 
A standard expression for the dielectric function is the random phase approximation (RPA) where the polarization function is calculated in lowest order with respect to the interaction. In the static limit, the effective interaction (\ref{Veff}) gives the Debye result 
$V_{{\rm C}^{6+}, e}^{\rm Debye}(1,2,{\bf q},z)= V_{{\rm C}^{6+}, e}(q)/(1+\kappa^2/q^2)$
 with the Debye screening parameter $\kappa^2=\sum_i Z_i^2e^2 n_i/(\epsilon_0 k_BT)+\kappa^2_e$,
\begin{equation}
\label{kappa}
 \kappa^2_e=\frac{4 \pi}{k_BT} 2 \left(\frac{2 \pi \hbar^2}{m_e k_BT}\right)^{-3/2}
\frac{e^2}{4 \pi \epsilon_0}
\frac{1}{ \sqrt{\pi}} \int_0^\infty dt \frac{t^{-1/2}}{e^{t-\beta \mu_e}+1}
 =12 \pi^{5/2} \frac{e^2}{4 \pi \epsilon_0} n_e \beta \frac{ F_{-1/2}(\beta \mu_e)}{(\beta E_F)^{3/2}}.
\end{equation}
The expression for the Debye screening parameter $\kappa$ includes the contribution of free electrons ($\kappa_e$) which eventually become degenerate. Then, in contrast to the classical limit (\ref{Debye}), the electron contribution is given by a Fermi integral which, in the strongly degenerate limit, yields the Thomas-Fermi screening length instead of the Debye screening length, see Refs. \cite{OPAL,KKER,KSK}. 
The electron chemical potential $\mu_e$ is given by Eq. (\ref{nemu}).

For the dynamical self-energies $\Sigma_e(p,z), \,\,\Sigma_{{\rm C}^{6+}}(k,z)$ occurring 
in Eq. (\ref{BSE}), a systematic expansion is possible in terms of Feynman diagrams \cite{KKER}. 
For the electron quasiparticle shift $\Delta_e(p )$ (\ref{quasip})  the expansion $\Delta_e(p )=\Delta^{\rm Fock}_e(p )+\Delta^{\rm corr}_e(p )$ results.
As lowest order with respect to interaction, the Fock shift for the electrons
\begin{equation}
\label{HFe}
\Delta^{\rm Fock}_e(p )=\sum_q \frac{e^2}{\epsilon_0 q^2}f_e({\bf p}+{\bf q})
\end{equation}
is obtained (the Hartree term vanishes because of charge neutrality of the plasma). This shift is a typical quantum effect.
 Because the ions are treated classically under the conditions considered here, the corresponding 
contribution disappears. The further treatment of the electron contribution (\ref{HFe})
is postponed to the following Sec. \ref{sec:deg}.

We consider in this section the next order of the expansion of $\Delta_e(p )$,
the correlation shift (Montroll-Ward shift) $\Delta^{\rm corr}_e(p )$. It describes the formation of a screening cloud and has been intensely investigated.
In the so-called GW approximation, the RPA expression for the screened interaction can be used, and we find
the Debye shift
\begin{equation}
\Delta^{\rm corr}_e(p )= -\frac{\kappa e^2}{8 \pi \epsilon_0}
\end{equation}
in the  low-density, non-degenerate limit. 
Because this is a classical effect, describing the formation of the screening cloud as solution of the 
Poisson-Boltzmann equation, it applies also to the ions $a_i$ which are shifted according to the charge number $Z_i$,
\begin{equation}
\Delta^{\rm corr}_i(p )= -\frac{\kappa Z_i^2\,e^2}{8 \pi \epsilon_0},
\end{equation}
in particular $Z_6=6$ for ${\rm C}^{6+}$. With these expressions, the IPD in Debye approximation (\ref{Debye})
is found for the ionization/recombination reaction  $a_i +s\rightleftharpoons a_{i+1}+e+s'$ discussed above.

It is an advantage of the many-particle approach that systematic improvements can be given.
The correlation shift has the general form
\begin{eqnarray} \label{deltasigmaeps}
&& \repart \rmupper{\Sigma}{corr}{c}(p,\omega) = - {\cal P}\, \int \frac{d^3 {\bf q}}{(2\pi)^3}
 \int \frac{ d \omega'}{\pi} V_{cc}(q) 
\impart \varepsilon^{-1}(q,\omega'+i0)
 \frac{1+\rmnumb{n}{_B}(\omega')}{ \omega-\omega'-E_{c}({\bf p}+{\bf q})/\hbar}.
\end{eqnarray}
$ {\cal P}$ denotes the principal value, the index $c$ denotes 
electron as well as, for our system here,  the different carbon ions.
Instead of approximating the dielectric function by the RPA expression, which gives the Debye result,
we can use the fluctuation-dissipation theorem which relates the inverse dielectric function to the 
dynamical SF \cite{Lin17},
For a two-component plasma (free electrons with charge $-e$,  ions with effective charge $\bar Z e$ and charge 
neutrality $\bar Z n_i=n_e$),
the imaginary part of the inverse dielectric function can be expressed via the dynamical SFs, see also \cite{Chihara},
\begin{eqnarray}
\label{Imeps}
&& \impart \varepsilon^{-1}({\bf q},\omega+i0) 
 = \frac{e^2}{\varepsilon_0\, q^2} \frac{\pi}{\hbar\, \bks{1+\rmnumb{n}{_B}(\omega)}}
\left[\bar Z^2\,n_i S_{\rm ii}({\bf q},\omega) 
 -2 \bar Z \,\sqrt{n_e n_i}S_{\rm ei}({\bf q},\omega)+n_e S_{\rm ee}({\bf q},\omega) \right]\,.
\end{eqnarray}
After using a plasmon-pole approximation for $S({\bf q},\omega)$ \cite{Gregori,Lin17}, the dynamical response of the system is
determined by the plasmon pole frequency $\omega_{\rm pl}= (\sum_c e^2 Z_c^2 n_c/\epsilon_0 m_c)^{1/2}$.
Then, the integral over the frequency in (\ref{deltasigmaeps}) is executed.
Accounting for non-linear screening  \cite{Lin17},
 the ionic contribution to the single-particle shift is related to the static SF
\begin{equation}
\label{SFa}
\Delta_i^{\rm SF, ion-ion}=\frac{3(Z_i+1) e^2 \Gamma_i}{2 \pi^2 \epsilon_0 r_{\rm WS} \sqrt{(9 \pi/4)^{2/3}+3 \Gamma_i}}
\int_0^\infty \frac{dq}{q^2} S^{ZZ}_{\rm ii}(q),
\end{equation}
with $r_{\rm WS} = (4 \pi n_i)^{-1/3}$ and $\Gamma_i=Z_i^2 e^2/(4 \pi \epsilon_0 k_BT r_{\rm WS} )$.
We can use known expressions for the SF such as the approximation given in 
Ref. \cite{Struf} for recent calculations  \cite{Lin17}. As explained there, an improved description of experiments
\cite{Hoarty13,Vinko12,VCW14,Vinko15,Vinko18,ciricosta12,ciricosta16} has been obtained. For future work \cite{Linnew}, HNC calculations or DFT-MD simulations can be applied to implement the structure factor in expression (\ref{SFa}).

Instead of the phenomenological SP expression (\ref{SP}),  the IPD is related to the dynamical ion structure factor. 
Within the Green function approach, expression (\ref{SFa}) can be improved in a systematic way 
considering higher order diagrams. In particular, the electronic contribution to the correlated part 
of the self-energy shift $ \Delta^{\rm corr}_e(p ) $ can be improved. Expressions for the Montroll-Ward term
are found, e.g., in Ref. \cite{KKER}. 

In this work, we are not concerned with the improvement of
the correlated part of the self-energies that will be considered in a forthcoming work, see also  \cite{RKKKZ78,ZKKKR78,KKER,KSK,Lin17}, but focus on the effects of degeneracy. 
As discussed above, the Debye approximation for the correlation shift can be replaced by
the Stewart-Pyatt expression or more advanced approximations 
based on the dynamical structure factor if going to high densities.
We consider here the SP approximation (\ref{SP}) frequently used in IPD calculations,
to have a result of reference. Results for the corresponding IPD are shown in Fig. \ref{fig:2b} below,
where also the comparison with SF calculations  \cite{Lin17} is given.

\subsection{Degeneracy Effects}
\label{sec:deg}

To investigate the effects of degeneracy, i.e. Pauli blocking and Fock shifts, we simplify the  two-particle equation (\ref{BSE}).
The ions are considered as non-degenerate so that their contribution to the Pauli blocking term is dropped. 
In addition, we replace the dynamical screening by a statically screened (Debye) interaction $V_{{\rm C}^{6+}, e}^{\rm scr}({\bf q}) $ and introduce the quasiparticle shifts (\ref{quasip}).
\begin{eqnarray}
\label{BSE1}
&&\left[E_e(p )+\Delta_e(p )+E_{{\rm C}^{6+}}(k)+\Delta_{{\rm C}^{6+}}(k)
\right] \psi^{5+}_{\hat n}({\bf p},{\bf k})\nonumber\\&&
+[1-f_e(p )] \sum_{\bf q}V_{{\rm C}^{6+}, e}^{\rm scr}({\bf q}) \psi^{5+}_{\hat n}({\bf p}+{\bf q},{\bf k}-{\bf q})=
E^{5+}_{\hat n} \psi^{5+}_{\hat n}({\bf p},{\bf k})\,.
\end{eqnarray}

In adiabatic approximation, the  motion of electrons is separated from the motion of ions. More systematical, we introduce 
Jacobian coordinates, the center-of mass momentum $\bf P$ and the relative momentum ${\bf p}_{\rm rel}$. We use a separation ansatz 
for the wave function 
$ \psi^{5+}_{\hat n}({\bf p},{\bf k}) = \Phi_{\hat n}({\bf P})\phi_{\hat n}({\bf p}_{\rm rel})$. The center-of-mass motion is given by 
a plane wave. In limit  $m_e \ll M$ where ${\bf p}_{\rm rel}\approx {\bf p}$, we obtain for the relative motion
\begin{eqnarray}
\label{BSE11}
&&\left[E_e(p )+\Delta_e(p )\right] \phi_{\hat n}({\bf p})
+[1-f_e(p )] \sum_{\bf q}V_{{\rm C}^{6+}, e}^{\rm scr}({\bf q}) \phi_{\hat n}({\bf p}+{\bf q})=
E^{5+}_{\hat n,{\rm rel}} \phi_{\hat n}({\bf p})
\end{eqnarray}
so that $E^{5+}_{\hat n}=E_{{\rm C}^{6+}}(k)+\Delta_{{\rm C}^{6+}}(k)+E^{5+}_{\hat n,{\rm rel}}$.


The Fock shift of an electron with momentum $p$ is given by the expression (\ref{HFe}).
In the limit of strong degeneracy, $T \ll T_{\rm F}$, we approximate the Fermi distribution function 
as step function, $f_e(p )= \theta(p_{\rm F}-p)$. The Fermi wave number follows as 
$p_{\rm F}=(3 \pi^2 n_e)^{1/3}$.  
At zero temperature, we find for the Fock shift 
\begin{equation}
\label{HF00}
\Delta^{\rm Fock}_{e,T=0}(p)=-\frac{e^2}{4 \pi \epsilon_0} \frac{1}{\pi p} {\rm Re}\left[p\, p_{\rm F}+(p_{\rm F}^2-p^2)\arctan\left(\frac{p}{p_{\rm F}}\right)\right].
\end{equation}

The contribution of the Fock shift $\Delta^{\rm Fock}_{e}(0)$ to the shift of the continuum edge is given by the value at
$p=0$. At $T=0$ we have for the shift of the continuum edge
$ \Delta^{\rm Fock}_{e,T=0}(0)=
-\frac{2}{\pi} \frac{e^2}{4 \pi \epsilon_0} p_F$. 
At finite $T$, the Fock shift of the continuum edge, Eq.  (\ref{HFe}) at $p=0$, is calculated numerically.

Considering the bound states of the in-medium Schr{\"o}dinger equation (\ref{BSE1}), the ions $a_5$, two contributions arise owing to the electron degeneracy: 
the Fock shift (\ref{HFe}) which modifies the kinetic energy in the Schr{\"o}dinger equation as well as the Pauli blocking term 
in front of the interaction potential. Because the Fermi function occurring in both contributions depends on $T$ and $n_e$, the solution 
 $E^{5+}_{n}$  of the  Schr{\"o}dinger equation (\ref{BSE1}) also depends on these parameters. Both  contributions to the shift of $E^{5+}_{n}$, the Pauli blocking and the Fock shift, are  found
 solving the equation   (\ref{BSE11}) for the relative motion.
 
As example we give the shift $\Delta_0=\Delta_0^{\rm bound,\,Fock}+\Delta_0^{\rm bound,\,Pauli}$ 
of the ground state energy $E^{5+}_{0}$ in perturbation theory. 
The two-particle Schr{\"o}dinger equation without any medium corrections has the well-known hydrogen-like ground state solution 
$E^{(Z-1)+}_0=-Z^2 \frac{e^4}{(4 \pi \epsilon_0)^2} \frac{m_e}{2 \hbar^2}=- 13.602 \,Z^2\,\, {\rm eV}$ with $Z=6$ for the case considered here,
 and 
\begin{equation}
\label{phi0}
 \phi_0( p)= 8 \sqrt{\pi a_Z^3} \frac{1}{(1+a_Z^2 p^2)^2}, \qquad 
\psi_0(r)=\frac{1}{\sqrt{\pi a_Z^3}}e^{-r/a_Z},
\end{equation}
$a_Z=\frac{4 \pi \epsilon_0}{Ze^2} \frac{\hbar^2}{m_e}=a_B/Z$. 

The Fock shift $ \Delta_0^{\rm bound,\,Fock}$ of the bound state energy results in perturbation theory as 
average of the momentum-dependent Fock shift (\ref{HFe})
with this unperturbed wave function (\ref{phi0}),
\begin{equation}
\label{Fb}
 \Delta_0^{\rm bound,\,Fock}=-\sum_{p,q}\phi^2_0( p) \frac{e^2}{\epsilon_0 q^2} f_e({\bf p}+{\bf q})
=-\frac{32}{\pi } \int_0^\infty dp \frac{p^2 a_Z^3}{(1+a_Z^2 p^2)^4}\Delta^{\rm Fock}_e(p )\,.
\end{equation}
An explicit expression can be given for zero temperature ($T=0$)
 \begin{eqnarray}
 && \Delta_{0,T=0}^{\rm bound,\,Fock}=-\frac{e^2}{4 \pi \epsilon_0} \frac{2}{\pi a_Z} \frac{5 a_Z^3p_F^3+3 a_Z^5p_F^5}{(a_Z^2p_F^2+1)^2}\,.
      \end{eqnarray}
Compared to the Fock shift $\Delta^{\rm Fock}_{e}(0)$ of the continuum edge, the bound state Fock shift
$ \Delta_0^{\rm bound,\,Fock} $  is determined by the momentum-dependent Fock shift (\ref{HFe}). The latter becomes smaller near the Fermi momentum, so that the bound state Fock shift is also smaller compared to the  
Fock shift of the continuum edge.    
      
The Pauli blocking shift is given by
\begin{equation}
\label{Pb}
 \Delta_0^{\rm bound,\,Pauli}=-\sum_{p,q}\phi_0( p)f_e( p) V_{{\rm C}^{6+}, e}(q) \phi_0({\bf p}+{\bf q})
=\frac{Z e^2}{4 \pi \epsilon_0} \frac{4 a_Z^2}{\pi} \int_0^\infty f_e( p) \frac{p^2 dp}{(1+a_Z^2 p^2)^3}
\end{equation}
which becomes at $T=0$
\begin{equation}
 \Delta_{0,T=0}^{\rm bound,\,Pauli}=\frac{Z e^2}{4 \pi \epsilon_0} \frac{2}{\pi a_Z} 
\left[\frac{a_Z p_F (a_Z^2p_F^2-1)}{(a_Z^2p_F^2+1)^2}+\arctan(a_Z p_{\rm F})\right].
\end{equation}
At finite temperatures, the integrals in Eqs. (\ref{Fb}),  (\ref{Pb}) are calculated numerically.
 Note that both effects, the Fock shift and the Pauli shift, have different sign and compete partially.
 Calculations are shown in the following Sec. \ref{sec:results}, see Fig. \ref{fig:2b}.

\section{Results for carbon plasmas}
\label{sec:results}

We present results for the ionization degree of carbon plasmas in the WDM regime.
The composition of the carbon plasma for given mass density and temperature is determined by the abundances 
of ions C$^{i+}$ with different charge $Z_ie, Z_i=0,1, \dots, 6$ (including the neutral atom). The composition 
of the partially ionized plasma (PIP) is described by the partial densities (\ref{ni1}) obtained in Sec. \ref{QS}.

To start with we briefly recall the ideal PIP
model neglecting any medium effects. This approximation is applicable in the low-density region where we have nearly free
motion of the constituents of the PIP.
In this limiting case, we consider noninteracting ions C$^{i+}$ in its ground state and bound excited states. 
The electrons which are not bound to ions are considered as free electrons. 
The densities of the different components of the plasma are connected by the neutrality condition
$\sum_i Z_i n_i=n_e$. The ionization energies $I_i$ necessary to separate an electron from the carbon ion 
C$^{i+}$ are known ($I^{(0)}_0 = 11.2603$ eV, $I^{(0)}_1 = 24.3833$ eV, $I^{(0)}_2 = 47.8878$ eV,
$I^{(0)}_3 = 64.4939$ eV, $I^{(0)}_4 = 392.087$ eV, $I^{(0)}_5 = 489.9933$ eV).
In addition, the excited states must be included, data can be found in Ref. \cite{NIST}. 
To have convergent results, the Planck-Larkin expression (\ref{PLC5}) is used for the intrinsic partition function.
The solution of the Saha equations for the partial densities of different ions (ground state and excited state)
gives the average ionization degree $\bar Z$ and the corresponding free electron density $n_e=\bar Z n_C$ 
as function of the temperature $T$ and the density of carbon nuclei $n_C$ in the charge-neutral equilibrium state.
Results for $\bar Z$ are shown below in Fig. \ref{fig:2a} ("ideal mixture") for $T_1=100$ eV as function of $n_e$.
The convergent Planck-Larkin intrinsic partition functions are \cite{NIST} $\sigma_5^{\rm PL}(T_1)=266.241$,
 $\sigma_4^{\rm PL}(T_1)=54.197$, $\sigma_3^{\rm PL}(T_1)=1.6889$, $\sigma_2^{\rm PL
}(T_1)=1.2345$, $\sigma_1^{\rm PL}(T_1)=0.5459$, and $\sigma_0^{\rm PL}(T_1)=0.08885$.
For the ideal electron gas, the classical approximation  has been compared to the ideal Fermi gas (\ref{nemu}),
but effects of degeneracy are small in the region of density and temperature considered there.
However, an ideal, noninteracting plasma model with occasional reactions to establish chemical equilibrium
is not appropriate for a dense plasma where interactions have to be taken into account.

\begin{figure}[h]
  \centerline{\includegraphics[width=350pt,angle=0]{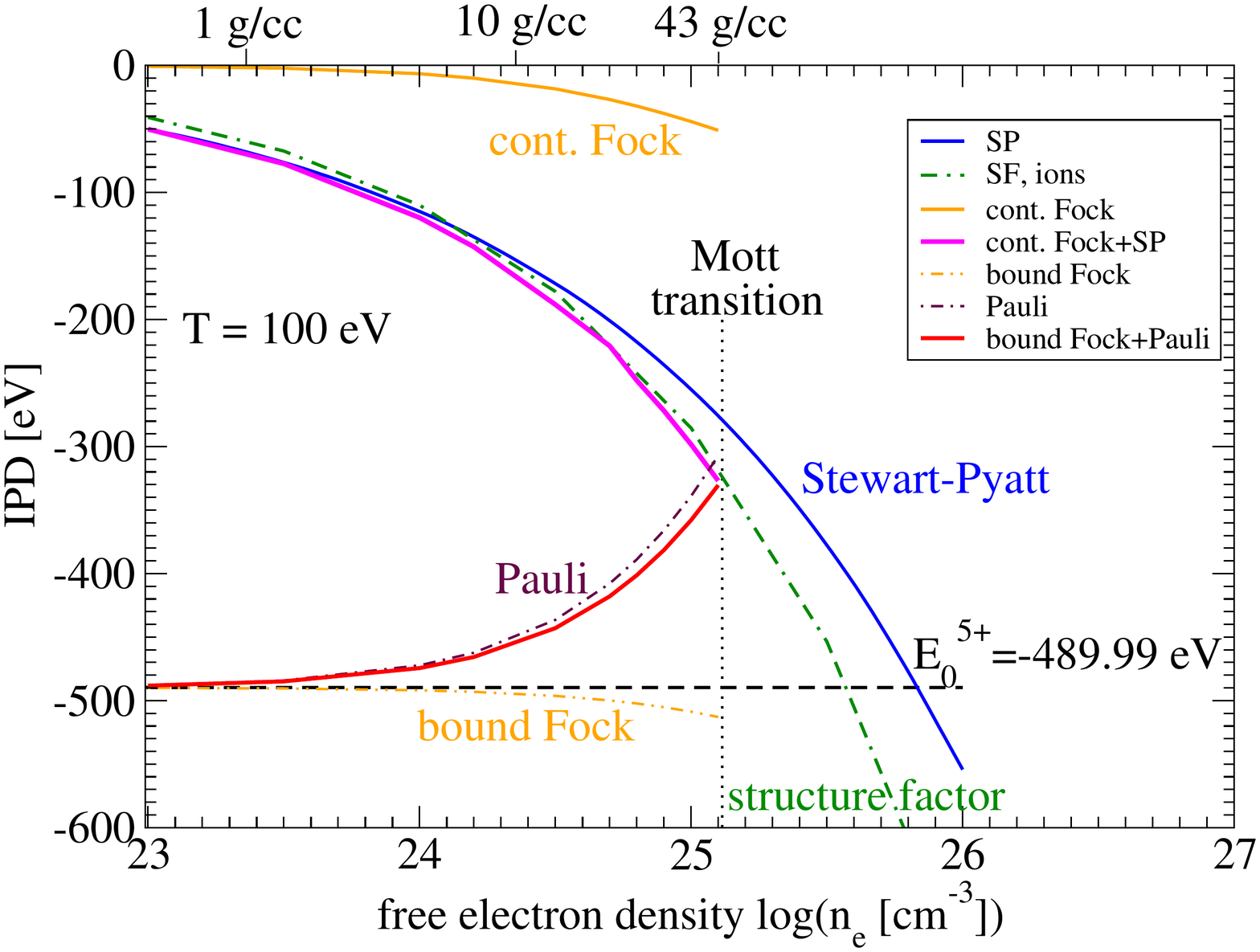}}
  \caption{Ionization potential depression (IPD) of C$^{5+}$ as function of the free electron density $n_e$
at fixed temperature $T = 100$ eV. The Fock shift of the continuum edge (cont.Fock), Eq. (\ref{HFe}) at $p=0$, 
together with the Stewart-Pyatt (SP) IPD (\ref{SP}) yields the shift of the continuum (cont.Fock+SP).
 For the ion C$^{5+}$ in its ground state, 
the Pauli shift  (\ref{Pb}) (Pauli) and the bound state Fock shift  (\ref{Fb}) (bound Fock)
as well as the sum of both (bound Fock+Pauli) are presented.
For comparison, the IPD obtained from the ionic structure 
factor shift of the continuum (SF,ions) as improvement of the SP model \cite{Lin17} is also shown.
The Mott transition is predicted at $n_e=1.3 \times 10^{25}$ cm$^{-3}$. Upper scale: Carbon mass density.}
\label{fig:2b}
\end{figure}

We now discuss the in-medium effects  such as 
Debye screening and its improvements by SP and SF as well as Pauli blocking, which determine the quasiparticle energies in the dense plasma for the hydrogen-like ion
C$^{5+}$. Results for the different contributions to the 
in-medium shifts are shown for $T=100$ eV as function of the free electron density $n_e$ in Fig. \ref{fig:2b}.
For the ground state of C$^{5+}$, the ionization potential in free space is $I_5^{(0)}= 489.9933$ eV. It is reduced by screening.
Improving the Debye result for the correlation part, the SP approximation (\ref{SP}) gives a Mott density 
$n_{e,\rm Mott}^{\rm SP}=6.89 \times 10^{25}$ cm$^{-3}$ where the bound state merges with the continuum.
Within the  quantum statistical approach \cite{Lin17} determined by the ionic structure factor ("SF,ions"), 
the IPD   
is larger, see Fig. \ref{fig:2b}. The corresponding Mott density for $T=100$ eV follows as  
$n_{e,\rm Mott}^{\rm SF}=3.78 \times 10^{25}$ cm$^{-3}$.

With increasing density, the effects of degeneracy of the electron subsystem become of increasing importance. 
Whereas the Fock shifts $\Delta^{\rm Fock}_e(0 )$ (\ref{HFe}) of the continuum edge and, even more, the Fock shift of the bound state (\ref{Fb}) 
remain small, the Pauli blocking (\ref{Pb}) becomes relevant for the dissolution of the bound state.
Taking into account all effects of degeneracy, the Mott density is further reduced
and the value $n_{e,\rm Mott}^{\rm deg}=1.28 \times 10^{25}$ cm$^{-3}$ is obtained. Note that the condition $\Theta=1$,
where the free electron system becomes degenerate, for a temperature of $T=100$ eV is satisfied at an 
electron density
$n_{e}^{\rm deg}=4 \times 10^{24}$~cm$^{-3}$, which corresponds to a mass density of 20~g~cm$^{-3}$ for a carbon plasma as mentioned in the Introduction.
Above this density, degeneracy effects, in particular Pauli blocking and Fock shifts, have to be considered.
Similar results are obtained also for the other ionization states of carbon.

In conclusion, the ionization potential $I_i = I_i^{(0)}+\Delta_i^{\rm corr}+\Delta_i^{\rm degen}$ contains 
 contributions due to correlations as well as degeneracy. 
At low densities the composition of the partially ionized plasma is well described using the 
IPD in Debye approximation or its improved versions, the semi-empirical SP or the quantum statistical SF
approaches, which also include strong correlation effects. The effects of degeneracy, in particular Pauli blocking, 
become of relevance in the region of higher densities where the free-electron system is degenerate, $\Theta \le 1$.
The region, where the plasma is nearly fully ionized, is strongly modified if Pauli blocking is 
taken into account.

\begin{figure}[h]
  \centerline{\includegraphics[width=350pt,angle=0]{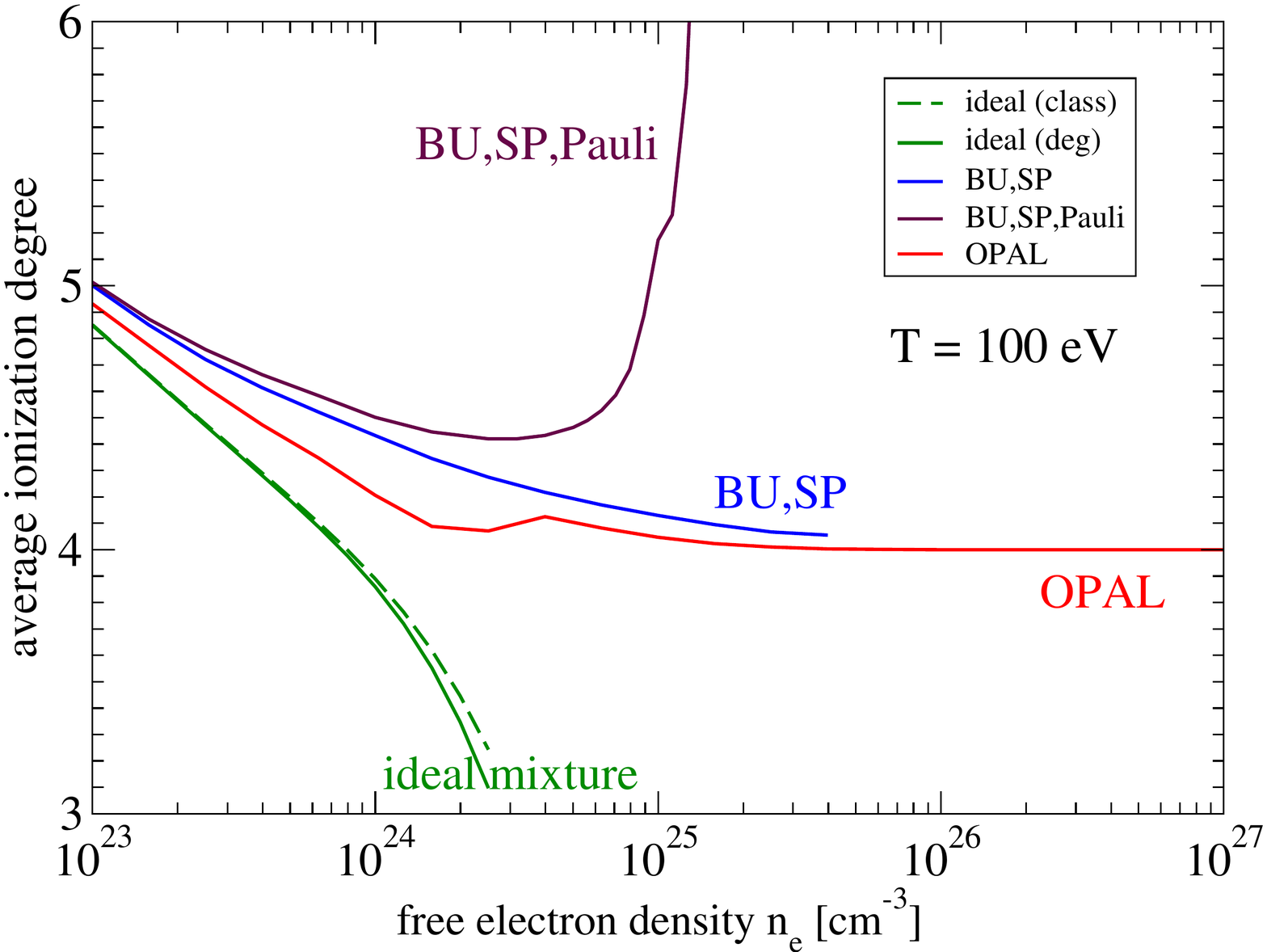}}
  \caption{Average ionization degree $\bar Z$ of carbon as function of the free electron 
density $n_e$ for temperature $T = 100$ eV. 
Ideal mixture with electron treated classically (ideal,class) and as Fermi gas (ideal, deg), 
OPAL and Stewart/Pyatt (BU,SP) are also shown. BU denotes the use of the Beth-Uhlenbeck expression 
(\ref{n2virialqu}), (\ref{virC5}) for the intrinsic partition function, SP the Stewart-Pyatt 
contribution Eq. (\ref{SP}). In addition the SP expression (\ref{SP}), the full IPD (BU,SP,Pauli)  
contains the Fock shift of the continuum (\ref{HFe}) at $p=0$ as well as the Fock shift of the bound state 
(\ref{Fb}) and the Pauli blocking (\ref{Pb}).}
\label{fig:2a}
\end{figure}

To demonstrate the effect of Pauli blocking on $\bar Z$, we performed calculations of 
the ionization degree of carbon
as function of the free electron density at a fixed temperature   $T=100$ eV, see Fig. \ref{fig:2a}.
Arbitrary ionization stages  $Z_i$ of carbon as well as excited states according to the NIST tables \cite{NIST}
have been included. 
 For the contributions of free and bound electrons to the density, we use the definition (\ref{n2virialqu}) 
but neglect the contribution of scattering states. The term -1 in the bound state contribution makes this part continuous near the Mott 
density where the bound state disappears. It compensates partly the contribution of scattering states according to the Levinson theorem.
Together with the extraction of the Born approximation, which is transferred to the quasiparticle shift \cite{ZS85,SRS}, we assume that the 
continuum contribution to the correlated density (second term of the right-hand-side of Eq. (\ref{n2virialqu})  becomes small and can be neglected. For further discussion see the conclusions.

Neglecting all in-medium effects, the approximation of an ideal mixture discussed above, becomes increasingly 
worse when $n_e$ exceeds
the value $10^{23}$ cm$^{-3}$. The account of IPD according to SP (denoted as "BU,SP" in   Fig.  \ref{fig:2a}) gives an ionization
degree of about $\bar Z = 4$ even at very high densities. This is also obtained from OPAL which is based on the SP
approximation for the IPD \cite{OPAL}. These values are used when measurements have been compared to theory 
\cite{Fletcher14,Kraus16}. More recent quantum statistical approaches \cite{Lin17} which relate the IPD to the ionic structure factor
give a slightly higher value for the ionization degree not shown here.

The ionization degree $\bar Z$ is found to be further increased if degeneracy effects, 
the Fock shift and the Pauli blocking, are taken into account. 
The corresponding ionization degree is shown in Fig.  \ref{fig:2a} (denoted as "BU,SP,Pauli"). 
The value $\bar Z =6$
appears for densities larger than the Mott density 
$n_{e,\rm Mott}^{\rm deg}=1.29 \times 10^{25}$~cm$^{-3}$, i.e. at a much lower density than predicted 
by the SP model.
The Mott effect predicts full ionization if all bound states merge with the continuum of 
delocalized electron states.
This is clearly seen using the virial form (\ref{n2virialqu}) of the intrinsic partition function, 
if only bound states are taken into account. 

The ionization degree of carbon at  $T=100$ eV has also been considered in Ref. \cite{Potekhin}. 
The calculations used an average 
atom model with different boundary conditions to mimic a band width, and the bound state contribution was 
defined  by the part of the band below the energy of the continuum edge. 
Qualitatively, the results are similar to the results 
for the ionization degree shown in Fig.  \ref{fig:2a}, denoted as "BU,SP,Pauli", and full ionization is predicted near the Mott
density. Similar calculations have been performed recently  
for lower temperatures and densities in Ref. \cite{Apfelbaum}. It is not clear to which extent correlation and degeneracy 
effects obtained from a systematic quantum statistical approach are already contained in those semi-empirical approaches.

\begin{figure}[h]
  \centerline{\includegraphics[width=400pt]{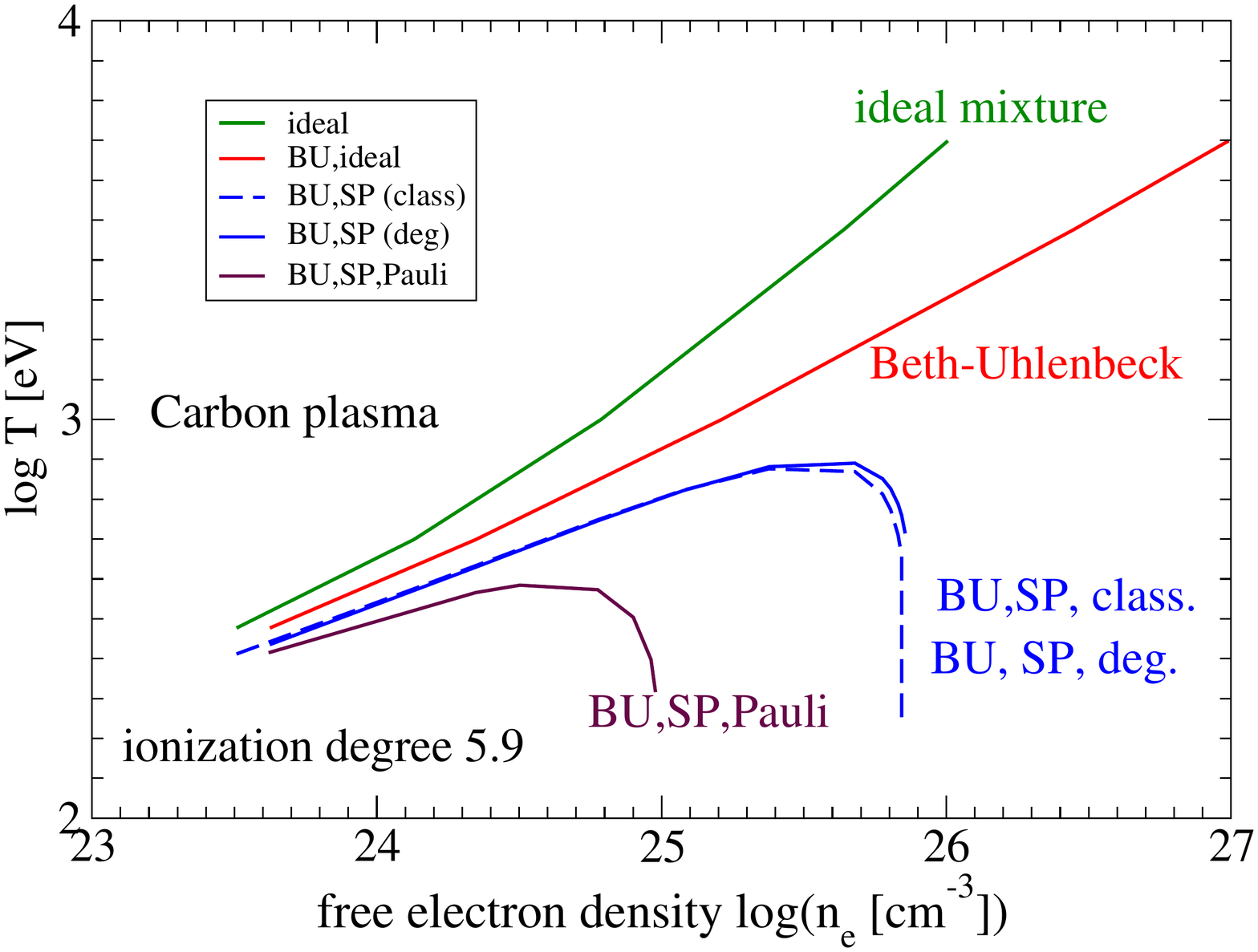}}
  \caption{Temperatures for the ionization degree of $\bar Z=5.9$ (iso-ionization lines) using different approximations for the IPD.
   $T^{\rm ideal}_{5.9}$ for the ideal mixture is compared with the ideal Beth-Uhlenbeck expression
$T^{\rm BU, ideal}_{5.9}$  (\ref{n2virialqu}), (\ref{virC5}), neglecting any medium effects. 
The account of the correlation shifts for the IPD, given by the Stewart-Pyatt model (\ref{SP}), 
yields the graph $T^{\rm BU,SP,class}_{5.9}$ for a classical free-electron gas or $T^{\rm BU,SP,deg}_{5.9}$
for the Fermi gas. In addition to the SP shift, the full IPD contains the Fock shift of the continuum (\ref{HFe}), $p=0$, as well as the Fock shift of the bound state (\ref{Fb}) and the Pauli blocking (\ref{Pb}). It gives the graph $T^{\rm BU,SP,Pauli}_{5.9}$.}
\label{fig:1}
\end{figure}

Of interest are the properties of WDM at high densities where the plasma becomes highly ionized.
Because there is no sharp transition to the fully ionized plasma, we consider the value $\bar Z = 5.9$ for the ionization degree as
a nearly fully ionized carbon plasma with only 10 percent hydrogen-like carbon ions. This concentration is decreasing with increasing temperature.
In Fig. \ref{fig:1}, we show
graphs of constant ionization degree $\bar Z=5.9$ (iso-ionization line) in the phase diagram $T, n_C$ (or $T,n_e$ with the relation $n_e=5.9\, n_C$) for which the plasma is nearly fully ionized.
Iso-ionization lines with $\bar Z=5.9$ in carbon are calculated for different approximations.

For an ideal mixture of non-interacting components in chemical equilibrium, treating the electrons classically,
the temperature $T^{\rm ideal}_{\bar Z=5.9}(n_e)$ increases with increasing density.
This behavior is only slightly shifted to lower temperatures if the IPD according to SP (\ref{SP}) is included.
Two corrections can immediately be done: the quantum description of the electron gas according to Eq. (\ref{nemu})
which determines the relation between density and chemical potential necessary for the chemical equilibrium (\ref{MAL}), 
and the term $-1$ occurring in the Beth-Uhlenbeck expression (\ref{n2virialqu}). Neglecting the IPD, the bound states are not dissolved
with increasing density. As compared to the "ideal" curve  $T^{\rm ideal}_{5.9}$, the more consistent $T^{\rm BU,ideal}_{5.9}$ which contains the Beth-Uhlenbeck form (\ref{virC5}) of the intrinsic partition function, is shifted to lower values, but also increases monotonically with density.

According to Eq. (\ref{virC5}), only bound states are taken into account, i.e.,
the ionization potential of the bound state must be positive. 
The Mott effect becomes visible if the IPD compensates the vacuum ionization potential $I^{(0)}_i$. 
The corresponding $T^{\rm BU,SP,deg.}_{5.9}$ in Fig. \ref{fig:1} 
using the Stewart-Pyatt approximation for the IPD
 shows a strong deviation from the other curves. (Note that the classical result for the electron chemical potential used for $T^{\rm BU,SP,class.}_{5.9}$ gives only small deviations.)
In particular, for electron densities higher than the Mott density $n_{e,\rm Mott}^{\rm SP}$
given above, all electrons are free, and the iso-ionization curve for $T^{\rm BU,SP,deg}_{5.9}$ 
abruptly goes to zero.
Even larger is the effect if exchange terms, in particular Pauli blocking, are included. The curve
$T^{\rm BU,SP,Pauli}_{5.9}$ in Fig. \ref{fig:1} indicates that the region of full ionization is reached already
at the Mott density $n_{e,\rm Mott}^{\rm deg}$ consistent with the results shown in Fig. \ref{fig:2b}.

The strong influence of the IPD on the onset of fully ionization, where $T_{5.9}$ is shifted 
to lower temperatures if Pauli blocking is taken into account, leads to higher values for 
the average degree of ionization $\bar Z$.
Higher values of $\bar Z$ have been observed in experiments \cite{Fletcher14,Kraus16} 
when comparing to OPAL \cite{OPAL} which is based on the SP approximation for IPD, 
but neglects degeneracy effects such as bound-state Pauli blocking.
Experiments \cite{Fletcher14,Kraus16} with CH mixtures show higher ionization degrees in comparison 
to the prediction of OPAL. For instance, the mean charge $\bar Z=4.9$ was measured  at density 6.74 g/cm$^3$ 
and $T = 86$ eV  in Ref. \cite{Kraus16} which is higher than the prediction $\bar Z=4.18$ of 
OPAL based on the SP approach. 
This discrepancy is only partly resolved using the SF approach 
\cite{Lin17} for this mixture. The account of Pauli blocking leads to a further increase of the  ionization
degree. Experiments for pure carbon plasmas at very high densities are in preparation at the NIF.

\section{Conclusions}

The Pauli blocking has to be taken into account for extreme high-density WDM when the electrons are strongly degenerate. 
This exchange effect seems to be essential for the appearance of high ionization degrees as compared to standard 
approaches considering only screening effects, e.g., SP used for opacity tables like OPAL
\cite{Hoarty13,ciricosta12,ciricosta16,preston13,Fletcher14,Kraus16,Crowley14,Stransky16,Calisti15,CFT15}.  
Within the many-body approach described in the present work, 
further effects such as the polarization shift of the bound states can be considered, and the  
correlation (Montroll-Ward) contribution to the electron self-energy can be improved 
taking higher-order Feynman diagrams into account. In addition, 
the perturbative solution of the in-medium Schr\"odinger equation (\ref{BSE11}) is improved.

The concept of the ionization degree or plasma composition is a useful approach to PIP but has to be used 
with care, in particular with respect to the inclusion of scattering states. 
The ordinary chemical picture which considers the PIP as a mixture of different components, 
the free particles as well as the bound  clusters, neglects the correlations between these components. 
The so-called physical picture, where only the "elementary" constituents (electrons and nuclei) and their 
interaction are considered, provides a consistent description of WDM. 
The drawbacks of the ordinary chemical picture are avoided if  spectral functions are considered,
which are well defined at arbitrary densities. Single-quasiparticle states and bound states are approximations
for the spectral functions where the energy levels are shifted and broadened because of the interaction with the 
plasma environment. In particular, the broadening of energy levels (Inglis-Teller effect \cite{IT,Lin17a}) has
to be considered if the signatures of bound states as separate peaks in the spectra disappear.  

A challenge is 
the use of density-functional theory \cite{Vinko12,VCW14,Vinko15,Vinko18} 
where the single-particle density of states is evaluated. Assuming that the broadening of the bands 
is less important for the integral over the spectral function, see Eq. (\ref{specn}), 
the shifts of the bands can be compared with the level shifts in our approach.
 Work in this direction is in progress \cite{Mandy}, see also \cite{PNP16} where orbital-free 
molecular dynamics is performed, and the sensitivity of the equations of state, obtained there, to the choice 
of exchange-correlation functionals is investigated.  Correct results for thermodynamic quantities are also available from PIMC calculations \cite{Militzer} in the high-temperature region where the difficulties using a nodal structure are less relevant. 
Controversies such as the treatment of strongly degenerate 
 systems \cite{Hu,Rosmej} where $\mu_e \approx E_F$ may be resolved within the quantum statistical approach, considering the contribution of  scattering phase shifts, see \cite{Phaseshifts}. For the strongly degenerate electron gas, bound-state like contributions do not disappear  
 if the bound state merges with the continuum of scattering states. At zero temperature, correlations in the continuum give a contribution to the correlated density until the bound state merges with the Fermi energy. 

The full solution of the quantum statistical 
 approach, including the contribution of scattering states, is needed to obtain a consistent description of 
physical properties of the partially ionized plasma. This is possible in the "physical" picture, i.e. the 
solution of the many-body problem for interacting electrons and nuclei. The ordinary "chemical" picture is 
improved using the quasiparticle concept. Instead of free particles, single-quasiparticle states are introduced 
which contain already contributions of interaction in mean-field approximation. 
In addition, correlations are defined which contain not only the in-medium bound states, 
but also the correlations in the continuum. 

The Pauli blocking is a quantum effect based on the antisymmetrization of the many-electron wave function.
It is only approximately described by an empirical potential for the interaction of bound states. 
A consistent description is given within the physical picture, solving the few-particle 
in-medium Schr{\"o}dinger equation (so-called Bethe-Salpeter equation)  which contains 
the phase-space occupation in the interaction term. The expression for an uncorrelated medium (\ref{BSE1}) 
given by the Fermi distribution function should be improved taking correlations in the medium into account, see Refs. \cite{cmf}.
In conclusion, the Pauli blocking is essential to describe the dissolution of bound states and the increase of the ionization degree at high densities when the WDM is strongly degenerate.

\section{ACKNOWLEDGMENTS}
The work of T. D.  was performed under the auspices of the U.S. Department of Energy by Lawrence Livermore National Laboratory under Contract No. DE-AC52-07NA27344 and supported by Laboratory Directed Research and Development (LDRD) Grant No. 18-ERD-033. 
D.B. and G.R. acknowledge partial support from the National Research Nuclear University 
(MEPhI) in the framework of the Russian Academic Excellence Project under contract no.
02.a03.21.0005.
 D.B. was supported by the Russian Science Foundation under grant number 17-12-01427.
R.R acknowledges support from the DFG via the FOR 2440.

\end{document}